# Interfacing Graphene-Based Materials With Neural Cells

**Mattia Bramini,[1,2]ϕ  Giulio Alberini,[1,3]  Elisabetta Colombo,[1,2]  Martina Chiacchiaretta,[1,3] Mattia Lorenzo DiFrancesco,[1,2] José Fernando Maya-Vetencourt,[1] Luca Maragliano,[1] Fabio Benfenati,[1,2,3]* Fabrizia Cesca[1,2]ϕ***

[1] Center for Synaptic Neuroscience and Technology, Istituto Italiano di Tecnologia, Genova, Italy
[2] Graphene Labs, Istituto Italiano di Tecnologia, Genova, Italy
[3] Department of Experimental Medicine, Università degli Studi di Genova, Genova, Italy

**Correspondence:**
ϕ Corresponding Authors:
Mattia Bramini         mattia.bramini@iit.it
Fabrizia Cesca         fabrizia.cesca@iit.it

Center for Synaptic Neuroscience and Technology
Istituto Italiano di Tecnologia
Largo Rosanna Benzi, 10
16132 – Genova, Italy

* Senior authors

**Running title: Graphene for Neuroscience**

**Abstract**
The scientific community has witnessed an exponential increase in the applications of graphene and graphene-based materials in a wide range of fields, from engineering to electronics to biotechnologies and biomedical applications. For what concerns neuroscience, the interest raised by these materials is two-fold. On one side, nanosheets made of graphene or graphene derivatives (graphene oxide, or its reduced form) can be used as carriers for drug delivery. Here, an important aspect is to evaluate their toxicity, which strongly depends on flake composition, chemical functionalization and dimensions. On the other side, graphene can be exploited as a substrate for tissue engineering. In this case, conductivity is probably the most relevant amongst the various properties of the different graphene materials, as it may allow to instruct and interrogate neural networks, as well as to drive neural growth and differentiation, which holds a great potential in regenerative medicine. In this review, we try to give a comprehensive view of the accomplishments and new challenges of the field, as well as which in our view are the most exciting directions to take in the immediate future. These include the need to engineer multifunctional nanoparticles able to cross the blood-brain-barrier to reach neural cells, and to achieve on-demand delivery of specific drugs. We describe the state-of-the-art in the use of graphene materials to engineer three-dimensional scaffolds to drive neuronal growth and regeneration *in vivo*, and the possibility of using graphene as a component of hybrid composites/multi-layer organic electronics devices. Last but not least, we address the need of an accurate theoretical modeling of the interface between graphene and biological material, by modeling the interaction of graphene with proteins and cell membranes at the nanoscale, and describing the physical mechanism(s) of charge transfer by which the various graphene materials can influence the excitability and physiology of neural cells.

## 1. Introduction

Graphene (G) is a single- or few-layered sheet of $Sp^2$-bonded carbon atoms tightly packed in a two-dimensional (2D) honeycomb lattice, with a thickness of only 0.34 nm (Geim, 2009). Each carbon atom has three µ-bonds and an out-of-plane π-bond that can bind with neighboring atoms (Geim, 2009), making G the thinnest compound ever known at one atom thick and the strongest compound discovered. Moreover, it is light, flexible and transparent and both electrically and thermally highly conductive, which opens the possibility of using it in a broad spectrum of applications, including supercapacitors (Sahoo et al., 2015;Casaluci et al., 2016) (Hess et al., 2011), flexible electronics (Eda et al., 2008;Meric et al., 2008), printable inks (Zhu et al., 2015;Bonaccorso et al., 2016), batteries (Hassoun et al., 2014;Dufficy et al., 2015), optical and electrochemical sensors (Pumera, 2009;Du et al., 2010;Kang et al., 2010), energy storage (El-Kady and Kaner, 2013;Bonaccorso et al., 2015;Ambrosi and Pumera, 2016) and medicine (Novoselov et al., 2012;Casaluci et al., 2016;Kostarelos et al., 2017;Reina et al., 2017). G-related materials (GRMs) include single- and few-layered G (1-10 layers; GR), G oxide (single layer, 1:1 C/O ratio; GO), reduced G oxide (rGO), graphite nano- and micro-platelets (more than 10 layers, but <100 nm thickness and average lateral size in the order of the nm and µm, respectively), G and G oxide quantum dots (GQDs and GOQDs, respectively), and a variety of hybridized G nanocomposites (Bianco, 2013;Wick et al., 2014;Cheng, 2016). Having such different composition and structures, these compounds possess very diverse properties that have to be taken in consideration when planning biomedical applications, as they elicit completely different biological responses. Thus, it is fundamental to properly identify and characterize the GRMs employed, to overcome the widespread lack of reproducibility affecting biological experiments with G materials.

In the last few years, biomedical applications of G have attracted an ever-increasing interest, including the use of G and GRMs for bioelectrodes, bioimaging, drug/gene/peptide delivery, nanopore-based DNA-sequencing, stem cell differentiation and tissue engineering (Feng et al., 2013;Yang, 2013). Moreover, GRMs have generated great interests for the design of nanocarriers and nanoimaging tools, two- and three-dimensional tissue scaffolds, anti-bacterial coatings and biosensors (Bitounis et al., 2013;Ding et al., 2015). The interest in using GRMs in medicine lies chiefly upon the extraordinary properties of G, including its mechanical properties, flexibility, transparency, thermo-electrical conductivity and good biocompatibility. GRMs could therefore overcome the limitations of metals and silicon, which are currently used for implantable devices, but are characterized by elevated stiffness, high inflammatory potential and poor long-term stability in physiological environments. Moreover, the biomedical field witnesses a strong need for innovative therapies to assess the increasing demand of more specific, safer and effective treatments for pathological conditions. Given these premises, a large amount of research on G focuses on medical applications, and particularly in the field of neurology, where its mechanical and electronic features make it a strong candidate for replacing current devices (Kostarelos et al., 2017;Reina et al., 2017).

Another appealing aspect of GRM-based medical devices lies on the increasing evidences of G biocompatibility, an extreme important issue to take into consideration for any new biomaterial brought to the market. Due to its chemistry, G surface allows strong and non-destructive interactions at the cellular level, which could even be improved by specific chemical functionalization (Cheng, 2016;P. Kang, 2016). This is particularly true for G-based supports and scaffolds oriented to tissue repair and regeneration, and in fact promising results have already been shown for neural and bone tissue engineering (Cheng, 2016;Reina et al., 2017). For what concerns G nanosheet dispersions, mostly intended for drug/gene delivery and diagnostic imaging purposes, the scenario is instead more complex (Bramini et al., 2016;Mendonca et al., 2016a;Rauti et al., 2016). The safety of this material is indeed still a challenging problem to address and every case needs to be analyzed separately by taking into account the synthesis method, the quality of the final product including its purity and the eventual presence of trace contaminants, as well as the biological environment in which G is to be applied.

**Graphene applications in neuroscience**. The biomedical applications of G represent a field in continuous expansion. Traditional treatments for central nervous system (CNS) disorders present a number of challenges, thus, developing new tools that outperform the state of the art technologies for imaging, drug delivery, neuronal regeneration and electrical recording and sensing is one of the main goal of modern medicine and neuroscience (Baldrighi et al., 2016). Since the development of carbon-related materials, nanotechnology has strongly impacted a number of applications (**Figure 1**) including: drug, gene and protein delivery, to cross the blood-brain barrier (BBB) and reach compromised brain areas; neuro-regenerative techniques to restore cell-cell communication upon damage by interfacing two (2D) or three (3D) dimensional scaffolds with neural cells; highly specific and reliable diagnostic tools, for *in vivo* sensing of disease biomarkers by cell labeling and real-time monitoring of biological active molecules; and neuronal activity monitoring and modulation, by highly sensitive electrodes for recordings and G-based platforms for electrical local stimulation (Mattei and Rehman, 2014;John et al., 2015;Chen, 2017;Kostarelos et al., 2017;Reina et al., 2017).

In detail, researchers have already started exploring the use of G at the CNS for cell labeling and real-time live-cell monitoring (Wang et al., 2014;Zuccaro et al., 2015); delivery to the brain of molecules that are usually rejected by the BBB (Tonelli et al., 2015;Dong et al., 2016); G-based scaffolds for cell culture (Li et al., 2013a;Menaa et al., 2015;Defterali et al., 2016); and cell analysis based on G-electrodes (Medina-Sanchez et al., 2012;Li et al., 2015). In addition, interfacing G with neural cells was also proposed to be extremely advantageous for exploring their electrical behavior or facilitating neuronal regeneration by promoting controlled elongation of neuronal processes (Li et al., 2011;Tu et al., 2014;Fabbro et al., 2016). These applications open up new research lines in neuro-therapeutics, including neuro-oncology, neuro-imaging, neuro-regeneration, functional neuro-surgery and peripheral nerve surgery (Mattei and Rehman, 2014).

In this review we will focus on few aspects of GRM research that we deem of particular interest for future neuroscience applications, i.e. (i) G as nano-carrier for drug and gene delivery, (ii) G interaction with the blood-brain barrier (BBB), and (iii) G-based 2D and 3D composites for neural regeneration, stimulation and recording. A final (iv) chapter is dedicated to an overview of computational modeling approaches that can help biologists and medical scientists to better understand the molecular and cellular interaction of G with living systems.

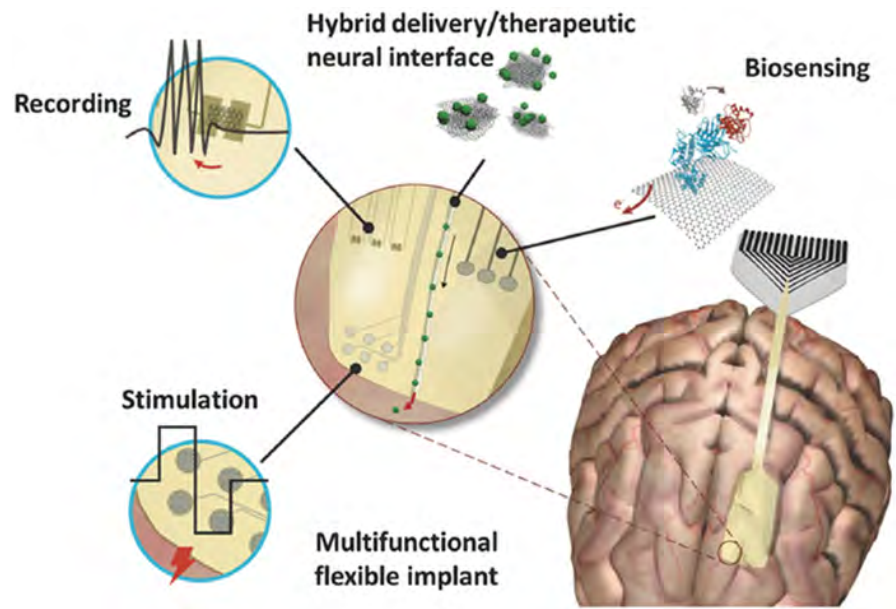

**Figure 1. Graphene based neural interfaces for a variety of neuronal functionalities like recording, stimulation and biosensing.** Modified with permission from Kostarelos et al. (Kostarelos et al., 2017).

## 2. How to reach the brain: G-based nanocarriers and the blood-brain barrier

### 2.1. Graphene nanosheet interaction with neural cells

Common mechanisms of cytotoxicity of G nanosheets have been reported in literature on different cell types, and include the physical interaction with cell membranes (Seabra et al., 2014); disruption of cell cytoskeleton (Tian et al., 2017); oxidative stress due to production of reactive oxygen species (ROS) (Chen et al., 2016;Mittal et al., 2016); mitochondrial damage (Pelin et al., 2017); DNA damage, such as chromosomal fragmentation, DNA strand breakages, point mutations, and oxidative DNA alterations (Fahmi et al., 2017); autophagy (Chen et al., 2014); and apoptosis and/or necrosis (Lim et al., 2016). Furthermore, published data suggest that GO is less toxic than G, rGO and hydrogenated-G; smaller nanosheets are less toxic than large flakes and highly dispersible G solutions are safer than aggregating ones (Donaldson et al., 2006;Bianco, 2013;Kurapati et al., 2016;Ou et al., 2016).

In the case of the CNS, the mechanisms of interaction of GRMs with neurons and astrocytes are still poorly investigated and unclear, depicting an undefined scenario mainly dependent on GRM intrinsic characteristics. Specifically, no changes in neuronal and glial cell viability were detected upon G exposure, both *in vivo* and *in vitro* (Bramini et al., 2016;Mendonca et al., 2016b;Rauti et al., 2016). However, primary neuronal cultures exposed to GO nanosheets displayed clear alterations in a number of physiological pathways, such as calcium and lipid homeostasis, synaptic connectivity and plasticity (Bramini et al., 2016;Rauti et al., 2016). Once internalized in cells, G nanosheets were seen to preferentially accumulate in lysosomes, as well as to physically damage mitochondria, endoplasmic reticulum and, in some cases, nuclei (John et al., 2015). Another study suggested that the irregular protrusions and sharp edges of the nanosheets could damage the plasma membrane, thus letting G entering the cell by piercing the phospholipid-bilayer (Li et al., 2013b). These features raise additional safety concerns, as free GRMs in the cytoplasm may lead to disruption of the cytoskeleton, impaired cell motility and blockade of the cell-cycle, similar to carbon nanotube-induced cytotoxicity.

The above-described effects were observed upon chronic G exposure, stressing the need of urgent and further biocompatibility assessment of the material with nerve tissues in long-term studies, hopefully linking *in vivo* effects to *in vitro* cellular and molecular interactions. A first strong evidence of G-induced CNS toxicity came from a recent *in vivo* study (Ren et al., 2016). To recreate a situation of G environmental pollution, researchers dispersed GO in water in the presence of *Danio rerio* (zebrafish) larvae. Exposed larvae displayed GO in the CNS and, most importantly, an induction of Parkinson's disease-like symptoms such as disturbance of locomotor activity, dopaminergic neurons loss and formation of Lewy bodies. These effects were likely a consequence of mitochondrial damage and apoptosis through the caspase 8 pathway, in the presence of a more general metabolic disturbance. G and GO nanosheets accumulate in small quantities in the CNS of rodents after intravenous (*i.v.*) injection without prior surface functionalization (Mendonca et al., 2016a;Mendonca et al., 2016b). rGO was also detected in brain tissues, particularly in the thalamus and hippocampus, after *i.v.* injection that was accompanied by blood-brain barrier (BBB) disruption (Mendonca et al., 2016b). Interestingly, rats treated with rGO flakes did not show any clinical signs of neurotoxicity, including no signs of tremor, convulsions, salivation, lacrimation, dyspnea and motor abnormalities. These findings are in contrast with the work carry out by Zhang *et al.* (Zhang et al., 2015), who reported the short-term decrease in locomotor activity and neuromuscular coordination in mice orally administered with rGO nanosheets. This discrepancy underlines that the route of administration is key parameter in determining G biocompatibility. Thus, the portal of entry of G into the organism, together with its dose, size, functionalization and aggregation, will determine the final biological effects.

In summary, the current data on G nanosheet biocompatibility are still controversial. This is due to the high heterogeneity of materials present on the market and the large variety of synthesis methods. Depending on the graphite source (starting material), the synthesis method, the use of chemicals and the dispersion form (solution or powder) of the final product, G can present different sizes, thickness, chemical surface and aggregation state, which all affect to various extent its interaction with the biological systems. It is clear, however, that G nanosheets may cause adverse environmental and health effects, leaving open the debate about their use as biomedical platform (Bramini et al., 2016;Reina et al., 2017). To date, GO nanosheets are preferred with respect to pristine G for biomedical studies, because of their major solubility and stability in biological fluids (Chowdhury et al., 2013;Servant et al., 2014a;Reina et al., 2017).

## 2.2. Graphene for biomolecule delivery to the central nervous system

As discussed above, the use of G nanosheet dispersions for biomedical applications could give some unwanted effects due to the intrinsic characteristics of the material. Interestingly, functionalizing the G surface could alleviate most of these drawbacks. The physical-chemical properties of G nanosheets can be tuned toward a higher degree of biocompatibility. Moreover, cargoes can be loaded via π-π stacking interactions, hydrogen bonding, or hydrophobic interactions (Georgakilas et al., 2016) giving the attractive possibility of using G as a platform for delivery of biomolecules that are usually rejected by the BBB. In fact, the large surface area available and the possibility of conjugating different molecules onto its surface, make G a suitable material for holding and carrying drugs, genes (including siRNA and miRNA), antibodies and proteins (Chen et al., 2013). In addition, it is also possible to modify its chemical structure by adding functional groups such as amino, carboxyl, hydroxyl, alkyl halogen, or azide groups (John et al., 2015). Surface functionalization has the double advantage of loading high quantity of biomolecules and specifically deliver them to target cells, while allowing a more homogenous dispersion of the material, since pure G is highly hydrophobic and tends to aggregate in aqueous solution, including biological fluids containing salts and proteins (Mattei and Rehman, 2014;John et al., 2015). Additionally, functionalized G nanosheets could be applied in systemic, targeted, and local delivery systems (Feng et al., 2011;Kim et al., 2011;Liu et al., 2013b). Thus, this approach could fulfill the increasing demand of multifunctional and versatile medical platforms.

Because of its unique fluorescent, photoacoustic and magnetic resonance profiles, several studies have also explored the possibility of incorporating G-based nanoparticles to enhance the *in vivo* visualization of brain tumors and improve tumor targeting of molecular anticancer strategies (Kim et al., 2011;Yang et al., 2012;Zhang, 2013;Hsieh et al., 2016). Also in this case, *in vivo* studies revealed that GO, more than GR, has good potential for these applications, in fact, systemically administered radiolabeled GO ($^{188}$Re-GO) could reach the brain parenchyma, although in a small amount (0.04%) (X. Zhang, 2011).

## 2.3 Blood-brain barrier crossing

The blood-brain barrier (BBB) is one of the most important physiological barriers in the organism, forming a dynamic interface that separates the brain from the circulatory system (Pardridge, 2001;Begley, 2004). The barrier is formed by cerebrovascular endothelial cells, surrounded by basal lamina and astrocyte perivascular endfeet that link the barrier system to the neurons (Abbott et al., 2010). Together with pericytes and microglial cells, endothelial cells support the barrier function and regulate its intercellular signaling to control the flow and trafficking to the brain (Dohgu et al., 2005;Abbott et al., 2010). The BBB, together with arachnoid and choroid plexus epithelium, restricts the passage of various chemical substance and foreign materials between the bloodstream and the neuronal tissue, while still allowing the passage of substances and nutrients essential to metabolic functions, from oxygen to various proteins, such as insulin and apolipoprotein E (Abbott et al., 2006;Strazielle and Ghersi-Egea, 2013). An interesting point is that the brain capillary endothelial cells clearly differ from the endothelial cells in the other districts of the body, in that they present a larger number of adherens and tight junctions between adjacent cells, so that no inter-cellular fenestrations exist (Abbott et al., 2006;Abbott et al., 2010). The tight junctions between the brain capillary endothelial cells are one of the most important structural and anatomical elements of the BBB. They create the major barrier, associating cell membranes tightly together and regulating paracellular movements of water, molecules, ions and other biomolecules (Begley and Brightman, 2003;Abbott et al., 2010). Based upon these characteristics, some researchers have highlighted that the permeability properties of the BBB reflect the tightness of intercellular junctions between brain capillary endothelial cells (Rubin et al., 1991). In other words, the low permeability characterising the BBB is caused for the most part by tight and adherens junctions that limit paracellular passage (Wolburg and Lippoldt, 2002). The result is that most molecular traffic is forced to take a transcellular route across the BBB, rather than moving paracellularly through the junctions, as in most endothelia (Abbott and Romero, 1996;Wolburg and Lippoldt, 2002;Hawkins et al., 2006).

To date, several mechanisms of transport across the BBB have been identified (**Figure 2**), including paracellular or transcellular pathways, transport proteins (carriers), receptor-mediated transcytosis and adsorptive transcytosis (Abbott et al., 2006). Transcytosis is a process whereby biomolecules are engulfed into a plasma membrane invagination and further transported from one side of the polarized cell monolayer to the other side. Specific proteins, such as insulin and transferrin, are taken up by receptor-mediated endocytosis and transcytosis, a process known as receptor-mediated transport (Kreuter et al., 2002;Rip et al., 2009;Ulbrich et al., 2009;Ulbrich et al., 2011). Native plasma proteins, such as albumin, are poorly transported, but cationisation can increase their uptake by adsorptive-mediated endocytosis and transcytosis (Abbott and Romero, 1996;Pardridge, 2007a). In addition to transcytosis, very small water-soluble compounds can penetrate the tight junctions through a paracellular aqueous pathway. In paracellular transport, tight junctions act as a "gatekeeper" and regulate paracellular diffusion of water-soluble agents. For example, sucrose is a water-soluble molecule and is able to cross the BBB in limited amounts by paracellular diffusion (Ek et al., 2006). The large lipid membrane surface area of the endothelium also offers an effective diffusive route (transcellular transport) for small gaseous molecules such as $O_2$ and lipid-soluble agents, including drugs such as barbiturates and ethanol. The endothelium furthermore contains transport proteins for glucose, amino acids, purine bases, nucleosides, choline and other substances. Some transporters, i.e. the P-glycoprotein, are energy-dependent and act as efflux transporters (active-efflux transport).

The complex network of transport systems described above, gives the BBB a vital neuroprotective function that however comes with some drawbacks, as the BBB also impedes the passage of drugs for CNS diseases. Pharmaceutical companies have invested significant effort and sums in trying to design drugs that can cross the BBB, with very limited success. It is reported that only 5% of the total amount of drugs developed for neuronal diseases actually reach the CNS (Pardridge, 2007b).

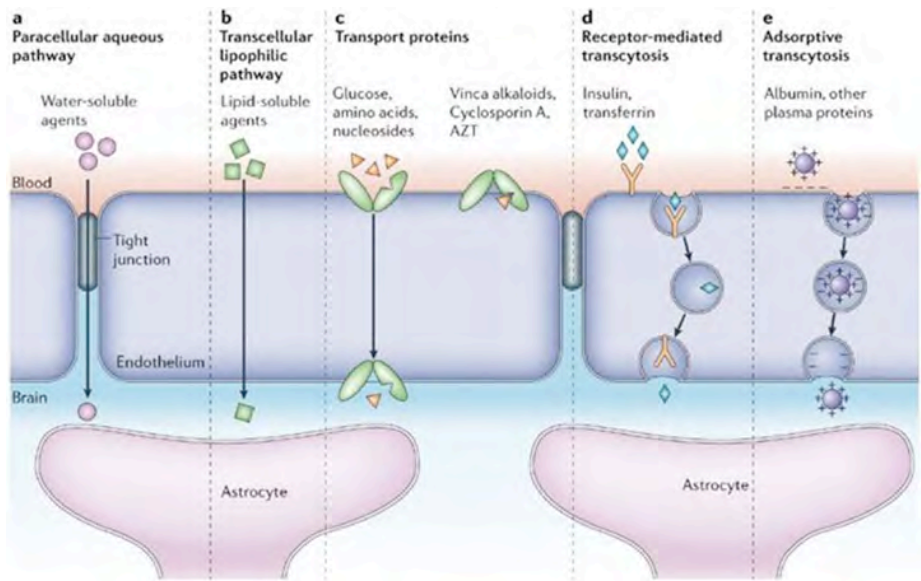

**Figure 2. Pathways across the blood-brain barrier.** Modified with permission from Abbott et al. (Abbott et al., 2006)

**2.4 Nanoparticle engineering**

The therapeutic potential of nanoparticles (NP) exposure depends chiefly on the rate of NP penetration when delivered from the external environment to the internal bio-compartments. Thus, biological barriers are central in determining the biological impact of NP exposure. Nanomaterials offer enormous potential for therapeutics and diagnosis, but also raise the possibility of unintended access to the brain (Herda, 2014). *In vivo* studies showed that nanoparticles could be found in the CNS upon various ways of administration (Semmler-Behnke et al., 2008;Zensi et al., 2009;Zensi et al., 2010). In parallel, *in vitro* models of human and murine BBB have been used and developed for the investigation of NP translocation (Andrieux and Couvreur, 2009;Ragnaill et al., 2011;Bramini et al., 2014;Herda, 2014;Raghnaill et al., 2014).

Numerous nano-delivery systems have been proposed and tested for therapeutic purposes, both *in vitro* and *in vivo* (Pandey et al., 2016). Amongst the state-of-the-art systems, polymeric NPs are promising because of their high drug encapsulation capacity, so they protect and transport hydrophobic drugs without damaging the BBB structure (Tosi et al., 2008). Binding apolipoprotein E to NPs has been suggested as a mechanism via which NPs could utilize existing pathways to access the brain (Kreuter et al., 2002;Wagner et al., 2012), and it indeed enhances the uptake of drugs (Michaelis et al., 2006). This approach is particularly promising with liposomes, which are highly biocompatible (Re et al., 2011). In general, exploiting receptor-mediated transcytosis by linking specific peptides to the NP surface has been the most studied system in the field of BBB crossing. Various molecules, such as transferrin, insulin, lectin and lipoproteins, physiologically use this route to pass from the blood stream to the brain; thus these ligands could increase the passage ratio of drug-loaded NPs through the BBB for therapeutic purposes (Herda, 2014;Åberg, 2016;Pandey et al., 2016). Recently, exogenous peptides known to undergo transcytosis in the BBB were also grafted to the NP surface to enhance their entrance in the CNS. Here, particular attention has been given to the diphtheria toxin receptor (DTR) and the HIV-TAT proteins. A mutant of DTR with no toxicity or immunogenicity has been tested to transport nano-liposomes and polybutylcyanoacrylate NPs across the BBB, both *in vitro* and *in vivo*, and indeed, only grafted NPs were able to transcytose the barrier (van Rooy et al., 2011;Kuo and Chung, 2012;Kuo and Liu, 2014). The same strategy was successful when using a derivate of the HIV-TAT protein, linked to the surface of polymeric micelles or $SiO_2$ NPs through PEG molecules (Liu et al., 2008a;Liu et al., 2008b;Zhao et al., 2016a). In addition, antibody-grafted NPs have been synthesized to specifically target brain vascular endothelium receptors (Loureiro et al., 2014;Saraiva et al., 2016), in order to exploit the physiological transcytosis mechanisms of the BBB. Once again, the most promising results were obtained with antibodies anti-Insulin (Ulbrich et al., 2011), anti-transferrin (Clark and Davis, 2015) and anti-LDL (Kreuter, 2014) receptors. Even though these recent developments in antibody engineering have improved the knowledge on brain therapeutics, by increasing targeting specificity and avoiding peripheral loss of the material, still significant efforts have to be made to translate these findings from research to clinical applications.

A major challenge of spherical NPs is that it is difficult to obtain a multifactorial engineered system able to encapsulate a drug, cross the BBB by receptor-mediated endocytosis and finally target a specific cellular subpopulation. In fact, even though NPs present a high surface area, the room for engineering peptides and molecules on the surface to drive and guide the system towards various body compartments is still limited. In this scenario, new approaches that combine external modulation of BBB permeability with NP engineering have been recently developed and are currently under investigation.

## 2.5   Surfactants coverage and hyperthermia

A very similar approach to the above-described NP surface modification with ligands consists in covering NPs with surfactants (Pardridge, 2012). This strategy induces a transient disruption of tight junctions, leading to higher permeability of the endothelium, thus allowing large molecule and nano-carriers to easily cross the BBB and reach the brain (Pardridge, 2012;Saraiva et al., 2016). Moreover, poly(sorbate 80) can adsorb apolipoprotein E and/or A-I, additionally giving NPs the capacity of binding lipoprotein receptors expressed in the brain endothelium and crossing the BBB (Kreuter et al., 2003;Petri et al., 2007).

In the last few years, researches have adopted innovative strategies with the aim of reduce BBB damage and increase the amount of drug transported to the CNS. One stream of research aimed at obtaining the time- and area-specific upregulation of BBB permeability, to facilitate NP passage. This was achieved, for example, by activating the A2A adenosine receptor, which increases the intercellular space between the brain capillary endothelium (Gao et al., 2014). A similar effect can be obtained by physically interacting with the BBB by inducing hyperthermia, a procedure that increases the local temperature of the endothelium to 41-43 °C. The change in temperature acts by selectively disrupting tight junctions and increases the paracellular permeability of the BBB. Interesting results were obtained using focused ultrasounds (FUS) and microbubbles, which showed very low tissue toxicity and high accumulation of doxorubicin in the CNS (Treat et al., 2007). Other techniques to produce hyperthermia are microwaves and radiofrequency. The latter has been tested *in vivo* for glioma treatment, in combination with classical chemotherapy and radiotherapy, displaying encouraging results (Wang et al., 2012). Finally, two more advanced strategies to induce hyperthermia have been recently tested: laser pulse and magnetic heating. Near-infrared ultrashort laser pulses induced BBB disruption in selected regions, thus allowing the passage of large molecules in the brain (Choi et al., 2011). Magnetic NPs (MNPs) were instead used for delivering bioactive compounds via heat generated from magnetic heating, using a low radiofrequency field (Tabatabaei et al., 2015). Since MNP location can also be monitored live, this technique can be applied for both treatment and diagnosis of diseases.

Despite the promising results, techniques that modulate and interfere with the BBB permeability, even if transiently and locally, are burdened by a major problem, i.e. that there is very poor control over the passage of unwanted molecules and/or microorganisms that populate the blood stream. If it is true that the amount of drug reaching the brain is increased upon tight junction opening, it is also true that toxic compounds, safely constrained to blood vessels by an intact BBB, may pass at the same time, posing high risks to the patient.

**2.6    Graphene and the BBB: a new way for drug & gene delivery to the brain.**

The key goal of any drug delivery system is to create a smart tool that recognizes specific targets and releases the drug in a controlled way (Allen and Cullis, 2004). The main limitation of G-based applications in neuroscience is its very low accumulation in the brain parenchyma upon intravenous injection. Once injected intravenously, G will engage with ions, lipids and proteins, resulting in the aggregation of the material and formation of a biomolecular corona that might affect the distribution of G and trigger inflammatory responses (Dell'Orco et al., 2010). In addition, nanosheets can be phagocytosed by macrophages, inducing activation and release of pro-inflammatory cytokines (Zhou et al., 2012), and interact with several blood components inducing hemolysis (Liao et al., 2011). Last but not least, G nanosheets could accumulate in the reticulo-endothelial system rather than in the tissue to which they are targeted (McCallion et al., 2016).

Particularly challenging is the passage through the BBB, which significantly limits the delivery of drugs, blocking roughly the 100% of large molecule neuro-therapeutics and more than 98% of all small-molecule drugs (Upadhyay, 2014). Accordingly to Mendonca et al., systemically injected rGO nanosheets cross the BBB through a transitory decrease in the BBB paracellular tightness and accumulate in the thalamus and hippocampus of rats (Mendonca et al., 2016b). On the contrary, functionalization of rGO with polyethylene glycol (PEG), usually used to improve biocompatibility of nanomaterials, induces BBB breakdown and astrocyte dysfunctions *in vivo* (Mendonca et al., 2016a). Among the various approaches to make G cross the BBB, ultrasounds were applied to mice to physically open BBB tight junctions and allow the drug delivery system to enter the brain. By following this method, GO nanosheets grafted with Gd-DTPA and poly(amidoamine) dendrimer, and loaded with EPI and the tumor suppressor miRNA *Let-7*, could reach the brain upon tail vein injection (Yang et al., 2014). The main advantage of this approach is the reversibility of the BBB opening. Interestingly, G allowed at the same time high contrast MRI analysis and a quantification of the distribution of the delivery system inside the brain tissue (Yang et al., 2014). These results are promising, however in-depth pharmacokinetics and toxicological studies are needed, especially for long-term treatments, keeping in mind that, with respect to what has been studied so far, this technique achieves a much higher G accumulation in the CNS.

Alternatively, G surface can be functionalized with specific biomolecules that enable the material to cross the BBB (Allen and Cullis, 2004;Goenka et al., 2014;John et al., 2015). A recent study has investigated an innovative nano-delivery system with high loading capacity and a pH dependent behavior. The GO@$Fe_3O_4$ nanocomposite was conjugated to lactoferrin (Lf), an iron transporting serum glycoprotein that binds to receptors overexpressed at the surface of vascular endothelial cells of the BBB and of glioma cells, in order to obtain Lf@GO@$Fe_3O_4$. After loading the NPs with doxorubicin (DOX), a drug used to treat glioma (**Figure 3**), NPs were intravenously injected and the particles were seen to migrate from the bloodstream to glioma cells (Liu et al., 2013a). NPs were more concentrated in the CNS compared to other organs, and a higher efficiency in tumor regression was observed, compared to the control of animals injected with DOX alone. Following a similar approach and with similar promising results, Yang *et al.* functionalized PEG-GO nanosheets with the Tat protein of the Human Immunodeficiency Virus (HIV), which allowed the drug-loaded PEG-GO system to cross the BBB by transcytosis, while leaving the barrier endothelium fully preserved (Yang et al., 2015a).

As previously discussed, another promising strategy to challenge the BBB is NP coating with surfactants (Kreuter et al., 2003;Gelperina et al., 2010). Kanakia *et al.* (Kanakia et al., 2014) improved GO delivery to the CNS by functionalizing the nanosheets with dextran; the material was found to cross the BBB and reach the brain without exerting toxic effects. Surprisingly, the GO concentration in the CNS increased with time, while remaining almost absent in other organs. Thus, the study suggests a slow accumulation of G in the CNS and long-term persistency of the material, that is encouraging from the point of view of the drug delivery system, but also raises safety concern on long-term toxicity of G nanosheets (Baldrighi et al., 2016), an issue that still needs to be assessed.

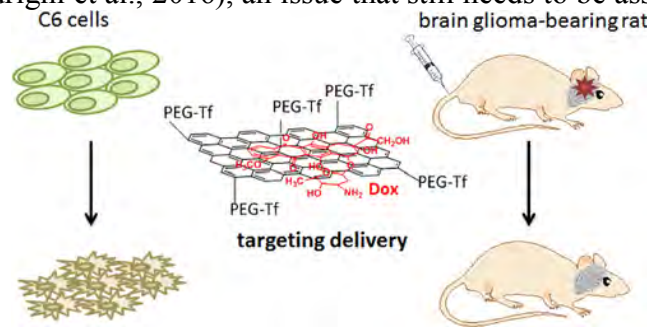

**Figure 3. Transferrin modified GO for glioma-targeted drug delivery.** Modified with permission from Liu et al. (Liu et al., 2013a)

The number of drugs successfully linked to G nanosheets is increasing. Liu and colleagues (Liu et al., 2008c) showed that GO-PEG flakes could be decorated with the water insoluble aromatic molecule 7-ethyl-10-hydroxy-camptothecin (SN38), via non-covalent van der Waals interactions. Similarly, other drugs, including different camptothecin analogues (Liu et al., 2008c), Iressa (gefitinib) (Liu et al., 2008c), and DOX (Sun et al., 2008), were successfully attached onto the GO-PEG complex by simple non-covalent binding. rGO-PEG particles were able to cross the endothelial layer of the BBB without disrupting the tight junctions, in both *in vitro* and *in vivo* studies (Mendonca et al., 2016a;Mendonca et al., 2016b). Recently, Xiao et al. used GQDs conjugated to a neuro-protective peptide. Once injected intravenously in a murine model of Alzheimer disease, they were able to increase learning and memory, dendritic spines formation and decrease pro-inflammatory cytokine levels (Xiao et al., 2016).

One of the main applications of G-based drug delivery systems is anticancer therapy, by linking G composites with chemotherapeutics. Given their strong optical absorbance in the near-infrared (NIR) region, G-based hybrid materials are also intensively studied for their promising applications in cancer phototherapy (Liu, 2011;Honigsmann, 2013) (Robinson et al., 2011;Yang et al., 2012). The rationale beyond this approach is to exploit the heat produced by the G accumulated in tumor regions upon NIR laser stimulation to kill cancer cells. This technique was successfully applied *in vitro* using U251 glioma cells (Markovic et al., 2011). Such experimental approaches are of special interest, as they might help overcoming the limitations imposed by the BBB (Abbott, 2013), and are very promising especially for the treatment of very resistant and aggressive tumors, such as the glioblastoma.

The intrinsic properties of G in the visible (VIS) and NIR range make it also an attractive tool for bio-imaging (Zhang, 2013) both *in vitro* and *in vivo* (Gollavelli and Ling, 2012). For example, aptamer-carboxyfluorescein/GO complexes were employed for intracellular monitoring and *in situ* molecular probing of specific clusters of living cells, such as artificially implanted tumors in mice. GO nanosheets were also used for photo-acoustic imaging, which relies on the acoustic response on heat expansion following optical energy absorption (Wang et al., 2010;Yang et al., 2010;Qian et al., 2012). Specifically for CNS applications, *in vivo* studies showed that intracranial administered PEG-GO and its derivatives can be imaged in the brain by two-photon microscopy (Qian et al., 2012). Through this imaging technique, a 3D distribution map can be reconstructed in the brain parenchyma due to the high tissue penetration of the fluorescence signal of PEG-GO composites. These promising results could lead to the use of G as a diagnostic tool for imaging brain cancerous lesions, especially if the material is engineered with biomolecules that specifically target tumorigenic cells. Furthermore, once the targeting is achieved, G properties can be optimized according to the specific application, i.e. the size and oxidation state might be changed to shift the emission wavelength from VIS to NIR, which has a deeper tissue penetration, thus improving the depth of the diagnostic imaging device. By combining the optical properties of G with other biodegradable and functional materials, it will be possible to create G-based composites and hybrids suitable for several live-imaging applications. So far, most of the tools have been tested *in vitro* on cancer cell lines, and *in* *vivo* for cancer detection and diagnosis, leaving unaddressed the possibility of using them to explore and image the CNS (Zhang, 2013;Cheng, 2016).

Similarly to drug delivery, also genetic engineering can exploit G properties and open new opportunities in biomedicine. The concept in this case would be to deliver nucleic acids, i.e. DNA or various types of RNA molecules, including miRNA and shRNA, to specific target cell populations, to restore physiological conditions (Cheng, 2016). The development of non-viral systems is of great importance for future medical approaches as G could allow overcoming some of the intrinsic limitations of viral systems, such as difficulties in accommodating long nucleic acids, batch-to-batch variations, elevated costs and the immunogenicity of viral vector systems (Kim et al., 2011;John et al., 2015). Different strategies have been developed, including the decoration with positively charged polymers (PEI, BPEI), dendrimers (PAMAM) and polysaccharides, which enhance gene transfection efficiency by promoting the interaction with the cell membrane (Liu et al., 2014;Paul et al., 2014). Being the technique of functionalization the same, both drugs and genes can be delivered simultaneously using G-based hybrid materials (Zhang et al., 2011). This would exhibit a synergic effect, as it would bring a significant enhancement of drug as well as transfection efficiency. On this line, G-nanosheets were functionalized with the cationic polymer PEI, a non-viral gene vector that forms strong electrostatic interactions with the negatively charged phosphate groups of both RNA and DNA (Feng et al., 2011). A step further was taken by Chen *et al.*, that used PEI-functionalized GO for gene delivery yielding a high transfection efficiency in the absence of any cytotoxic effect (Chen, 2011).

In summary, G-based delivery systems, when conveniently functionalized or associated with complementary technologies, represent promising candidates for both diagnostics (i.e. imaging) and therapeutics (i.e. drug and gene delivery) neuroscience applications. Moreover, in spite of few studies showing toxic effects of exposure the nervous system to bare G and rGO (Bramini et al., 2016;Mendonca et al., 2016b;Rauti et al., 2016), to date there is no solid evidence that functionalized-G is harmful to neuronal cells and the BBB. Since G-based technologies for biomedical applications are constantly and rapidly evolving, the near future may see the development of new safe and highly neurocompatible materials.

## 3 Graphene substrates as neuronal interfaces

Tissue engineering aims to restore the functionality of a disrupted tissue by interfacing it with suitable biomaterials. This is a fast-expanding field of research in need of innovative approaches to achieve highly biocompatibile, functional and low invasive implants for long-term applications. For what concerns the nervous system, active and dynamic implantable devices are extremely advantageous as they allow to simultaneously stimulate and record electrical activity of neural cells. Various types of implantable devices have been developed to be used as neural interfaces. Amongst these are deep brain stimulations implants (DBI) for the electrical stimulation of deep structures in the CNS, clinically used to treat dystonia and tremor in Parkinson's disease (Perlmutter and Mink, 2006), retinal and cochlear implants to electrically stimulate the surviving neurons in the presence of retinal degeneration or to convert external sounds in electrical impulses (Spelman, 2006;Picaud and Sahel, 2014), central and peripheral nervous system stimulators for motor rehabilitation after spinal cord lesions (Hatsopoulos and Donoghue, 2009), and intracranial electrodes to map brain electrical activity for diagnostic purposes (Chang, 2015).

The intrinsic properties of G can be exploited to design G-based devices for neuronal interfaces, as G can enhance the optical, electrical and mechanical properties of composite nanostructures. In general, fundamental requirements for a good neural implant are a good biocompatibility coupled to minimal inflammatory response, adequate signal-to-noise ratio if neural recordings are envisaged, and minimal invasiveness, preserving the integrity of the implanted tissue. Typically, G-based scaffolds can be classified according to their dimensionality, i.e. one-dimensional (fibers, ribbons or yarns), two-dimensional (papers, films) and three-dimensional (Cheng, 2016;Reina et al., 2017). The most common applications of G-based structures in nanomedicine include the engineering of scaffolds for *in vivo* neuronal regeneration, stimulation and recording, and for on-demand drug delivery (Cong et al., 2014;Cheng, 2016). For what concerns *in vivo* applications, the use of 2D devices is mostly limited to planar electrodes (Liu et al., 2016;Park et al., 2018). In fact, several G-based 2D devices have been engineered, but due to technical limitations, so far they have been tested mainly *in vitro* (for a comprehensive review of 2D, G-based substrates applied to neuronal cells see book chapter Bramini *et al.* (*in press*) (Bramini, 2018). For example, in the case of CVD-G, the limiting step is the transfer of the monolayer G onto the final substrate, a process that often creates contaminants and defects in the G structure. In addition, a suitable substrate that will interfere as less as possible with the chemical-physical characteristics of G is still to be found. Furthermore, 2D devices were less active *in vitro* with neuronal stem cells compared to 3D scaffolds with the same surface chemistry (Jiang et al., 2016), clearly indicating that morphology, dimensionality, accessibility and porosity are critical scaffold features. Indeed, foams and hydrogels are the scaffolds of choice to drive regeneration in the brain, while directional conduits are preferred to drive re-growth of peripheral nerves. In the next paragraph we will discuss the latest developments in the use of 3D G-based scaffold in neuroscience, focusing on the link between the G content and structure of the device, and its functionality.

## 3.1 3D G-based scaffolds: composites, foams, fibers and hydrogels.

Applications of G-based materials in the neurology field will only be possible upon development of three-dimensional scaffolds able to support nerve regeneration across the injured/lesioned site. The unique properties of planar 2D G-scaffolds are exceeded by 3D G-structures, which provide a microenvironment where cells are able to grow under conditions that are closer to the *in vivo* situation. In addition, as previously mentioned, 3D structures possess an enormous interface area and provide highly conductive pathways for charge transport, useful to support neural network formation and neuronal regeneration.

Several 3D scaffolds have been generated and tested *in vitro*, however so far only a very limited number of them have also been implanted *in vivo* (**Figure 4**). Some examples include G-coated electrospun PCL microfiber scaffolds, which were implanted in the striatum or the subventricular zone of adult rats. G-coated implants were associated with a lower microglia/macrophage infiltration when compared to bare scaffolds, while supporting astrocytes and neuroblast migration from the SVZ (Zhou et al., 2016). Free-standing 3D GO porous scaffolds were implanted in the injured rat spinal cord, showing no local or systemic toxicity and a good biocompatibility also in the case of chronic implantation (Lopez-Dolado et al., 2015). Of note, long-term (30 days) implants were able to promote angiogenesis and partial axonal regeneration (Lopez-Dolado et al., 2016). No attempts have been made so far to use G-based materials to drive peripheral nervous system regeneration. A first step in this direction is represented by the engineering of G-silk fibroin composite nanofiber membranes. This composite material is of interest as it combines the electrical conductivity and mechanical strength of G with the good compatibility of silk. Although they have not been tested *in vivo*, G-silk membranes support the growth of Schwann cells *in vitro* (Zhao et al., 2017).

Various three-dimensional G and GO foams were shown to be compatible substrates for stem cells (Crowder et al., 2013;Li et al., 2013a;Serrano et al., 2014;Guo et al., 2016;Sayyar et al., 2016). Li et al. firstly described 3D G-based foams (3D-GFs) as suitable scaffold for neural stem cells (NSCs) growth and proliferation. NSCs grown on 3D-GFs were able to differentiate into neurons and astrocytes; moreover, it was also noticed that 3D-GFs were optimal platforms for electrical stimulation of NSCs in order to enhance their differentiation (Li et al., 2013a). Similar results have been obtained more recently with rGO microfibers, which could support NSC viability and drive them toward a neuronal phenotype (Guo et al., 2017). Interestingly, the features of the G scaffolds (i.e. stiff *vs* soft) differentially affected cell adhesion and proliferation and could drive neural stem cell differentiation toward the astrocyte and neuronal lineages, respectively (Ma et al., 2016). Hippocampal neurons cultured on 3D-GFs are characterized by a more extensive connectivity associated to a higher network synchronization with respect to 2D-G substrates, thus better mimicking the physiological properties of the brain (Ulloa Severino et al., 2016). Microglial cells were also grown on 3D G foams. In this case, the 3D structure of the scaffolds affected the neuroinflammatory response of the cultured cells, probably because of spatial constraints due to the 3D topographic features (Song et al., 2014). Similar to what described for 2D materials, also 3D G/GO scaffolds were used as cell stimulating electrodes, to drive neuronal growth and differentiation of NSCs (Li et al., 2013a;Akhavan et al., 2016).

A new generation of electro responsive 3D-G scaffolds is also being developed, i.e. G-based hydrogels, which mimic soft tissue and have been proposed for controlled, stimulation-triggered drug release applications. Hybrid G-based hydrogels are synthetized mainly using GO, G oxide peroxide (GOP) or rGO, by incorporating very low amounts of the material into a hydrogel matrix, to enhance its electrical, mechanical and thermal properties (Servant et al., 2014b). Such materials are able to support neuronal growth and the development of synaptic activity (Martin et al., 2017). Following a similar approach, dexamethasone, a corticosteroid medication, was loaded onto poly(lactic-co-glycolic) acid NPs that were subsequently added into alginate hydrogels. The final composite was used as coating of gold and iridium electrodes for local drug administration after implantation (Kim et al., 2004;Kim and Martin, 2006). These or similar strategies could be employed to engineer smart coating for neuronal implants, with the final goal of having a device able to release biologically active molecules upon controlled electrical stimuli, at the same time improving the surface softness and enhancing the biocompatibility of the implants.

Altogether, the use of G materials in 3D implants aimed at neuroscience applications is still limited. However, much is to be learnt from other fields of biomedicine. For example, G-hydrogels and foams have recently been proposed for anticancer therapy (Xu et al., 2017;Zhang et al., 2017), as well as for guided bone (Lu et al., 2016), cartilage (A. Nieto, 2015) and muscle (Mahmoudifard et al., 2016) regeneration. We expect that the cross-fertilization between these different disciplines will lead in the close future to the development of functional 3D, G-based implants for nervous system applications.

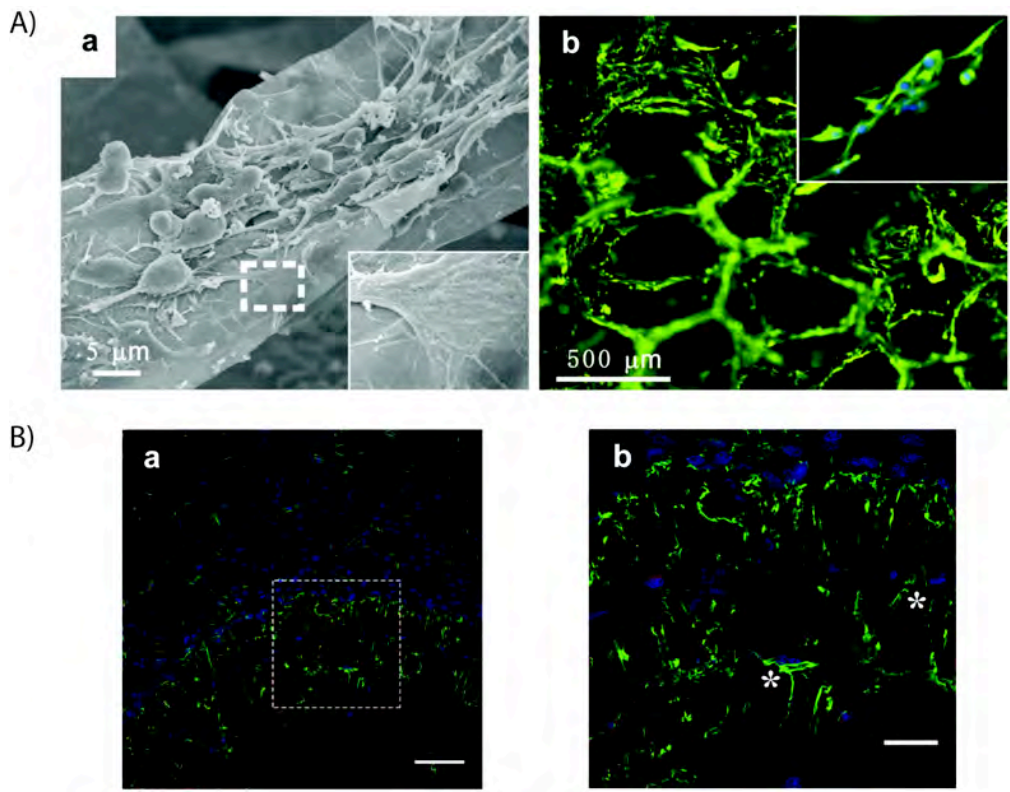

**Figure 4. 3D G-scaffolds *in vitro* and *in vivo*.** (A) (a) SEM images of NSCs cultured on 3D-G foams under proliferation conditions. The insets illustrate the interaction between the cell filopodia and surface. (b) Fluorescence images of NSCs cultured on 3D-G foams for 5

days. Nestin (green) is a marker for neural stem cells, and DAPI (blue) identifies nuclei. Modified with permission from Li *et al.* (Li et al., 2013a). (B) (a,b) Brain astrocyte/G-scaffolds interaction and astrocyte process infiltration 3 weeks after scaffold implantation. Green: GFAP-positive astrocytes, blue: DAPIstained nucleus, red: surface-functionalized scaffolds. (b) Detailed astrocyte morphology of the dash-box indicated area in (a). * indicate astrocytes that bridge a gap between two scaffold layers. Scale bar, 50 (a) and 20 (b) μm. Modified with permission from Zhou *et al.*(Zhou et al., 2016)

## 3.2 Graphene-based devices for neural recording and stimulation.

Clinical interventions for the recovery of neural dysfunctions and motor disorders attract and challenge the research toward implantable stimulation devices able to adapt to flexible supports and possibly outperform the most common metal electrode-based technologies. Polymeric interfaces outperform in terms of mechanically compliant properties, but often lack durability under physiological conditions, and, above all, proper electrical conductivity. Most of the neural stimulation performed so far with G-based electrodes in contact with living neuronal tissues or cells has been limited to modulate their growth and/or differentiation (Thompson et al., 2015).

Neural stimulation techniques, such as deep brain or cortical stimulation, cochlear and retinal implants, usually rely on the ability of the implanted devices to elicit a functional response of the tissue by providing minimum injected charge, and therefore require electrodes (Kostarelos et al., 2017). To date, *in vivo* studies show that G electrodes can stimulate and record neuronal activity. G electrodes produce slightly higher values of charge injection with respect to common noble metal electrodes, like Pt or Au. New promising materials and compounds exploit G to reach up to tens of mC cm$^{-2}$ charge injection levels, like in the case of an *in vivo* probe of laser reduced GO embedded into parylene-C (N. V. Apollo, 2015). The authors employ the novel flexible freestanding electrodes both to stimulate retinal ganglion cells *ex vivo* as much as to record neural activity *in vivo* from cat visual cortex. This constitutes one of the few reported evidences of neural stimulation with a G-based device. Other interesting applications make use of copper microwires encapsulated with CVD-G for an MRI compatible neural device (Zhao et al., 2016b), or flexible G micro-transistors for the mapping of brain activity (Blaschke, 2016) just to mention a few, but still limited to the recording of neural activity *in vivo*. In addition, Kuzum and colleagues (Kuzum et al., 2014) also developed a flexible, low noise G electrode for simultaneous electrophysiology and imaging recording *in vivo*. After bicuculline injection to evoke epileptiform activity, it was possible to register simultaneously from rat cortical hemispheres with G electrodes and Au electrodes of the same size. The G electrodes showed 6 times lower SNR with respect to Au electrodes, suggesting that the adoption of the new G-based recording system could offer clear advantages for studying brain electrical activity. In addition, thanks to the transparency of the G electrodes, it was also possible to image the cortical area, combining *in vivo* two-photon imaging and cortical electrophysiological recording (Kuzum et al., 2014).

A further progress has been achieved by developing G field-effect transistors (G-FETs), which allow signal amplification reducing external noise (Veliev et al., 2017). Flexible G-based supercapacitors showed recently their potential for neural stimulation thanks to their improved double layer capacitance when hybridized with polymeric materials, like PEDOT:PSS and rGO, G-polyaniline nanocomposites or CVD GO foams (Yang et al., 2015b;Hu et al., 2016). Nevertheless, most of these efforts need yet to be translated into usable electrodes for neuroscience applications. Another way to exploit G in bio-medical applications has been to enhance the optoelectronic properties of photosensitive neural interfaces deputed to the recovery of compromised vision. A polyimide array of photodetectors based on $MoS_2$ and inkjet G has recently been proposed as flexible retinal prosthesis, and tested for biocompatibility *in vitro* (Hossain, 2017).

In this framework, extensive reports demonstrate the ability of G to enhance organic photovoltaic devices. Diverse strategies have been followed to realize G-based solar cells, ranging from the modification of the anode electrode with GO (Rafique et al., 2017), to the realization of novel polymeric compounds containing GO-flakes or G quantum dots to improve charge carrier mobility or charge separation (Novak et al., 2016;Ye, 2016). These results depict a promising pathway to exploit the G electrical properties for biological applications.

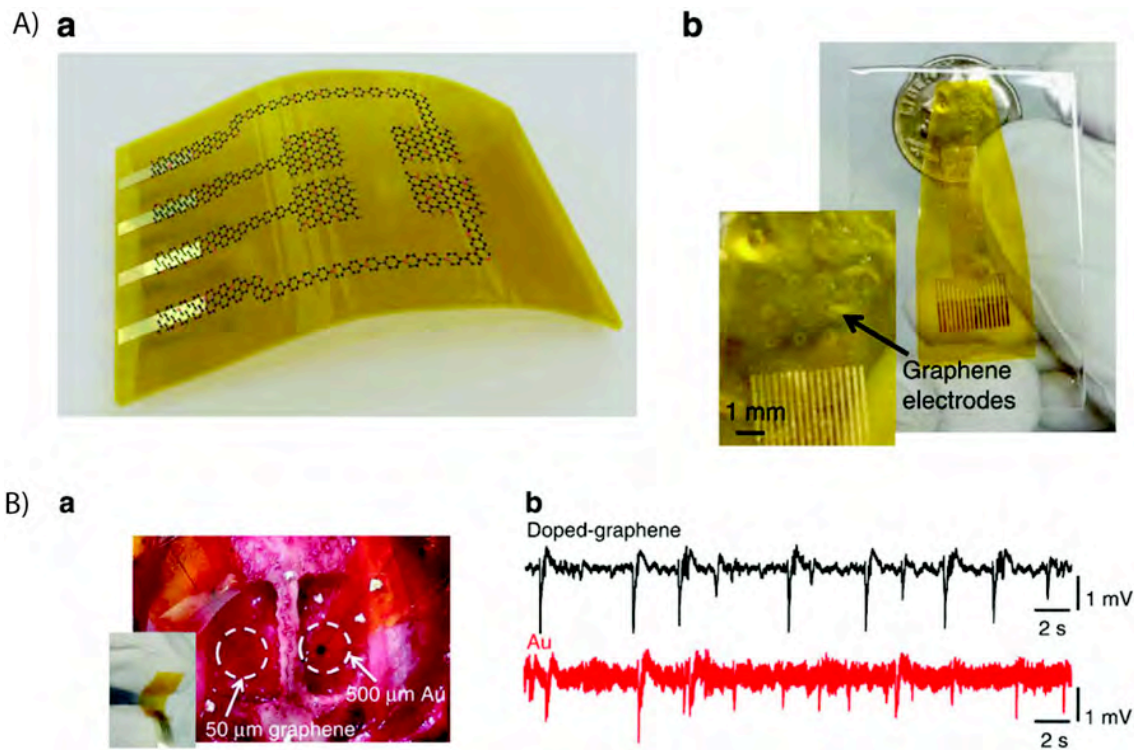

**Figure 5. Graphene electrodes for *in vivo* recording.** (A) Schematic illustration of a flexible G neural electrode array. (b) Photograph of a 16-electrode transparent array. The electrode size is $300 \times 300\ \mu m^2$. (B) (a) Photograph of a $50 \times 50\ \mu m^2$ single-G electrode placed on the cortical surface of the left hemisphere and a $500 \times 500\ \mu m^2$ single-Au electrode placed on the cortical surface of the right hemisphere. (b) Interictal-like spiking activity recorded by $50 \times 50\ \mu m^2$ doped-G and Au electrodes. Recordings with doped-G electrodes are five- to sixfold less noisy compared with the ones with same size Au electrode. Modified with permission from Kuzum *et al.* (Kuzum et al., 2014)

# 4. Computational modeling and simulations of Graphene-interacting biomolecular systems

Understanding the fine structural details underlying interactions between biomolecules and inorganic surfaces is pivotal for many applications in nanomedicine. Although relevant experimental results about the dynamics of these interactions have been recently reported, many topological details remain unclear, especially at the initial events at ns to µs timescales. To fill this gap, the use of computational modeling and Molecular Dynamics (MD) simulations gives a relevant contribution, providing details that cannot be accessed by experimental techniques (Ozboyaci et al., 2016).

With its promising properties, G has shown great potential in various applications, and the number of computational studies devoted to it is in constant growth (Cavallucci, 2016). Classical MD simulations (i.e. based on a classical physics description of atom-atom interactions) produced in the recent years a large amount of results on the interaction between G-based materials and biomolecules. In particular, these studies allowed to deeply characterize G as a substrate or nanopore for the deposition of biomolecules, differentiating the behavior of pristine G from that of GO. Moreover, MD simulations have been widely used to test G biocompatibility by studying its interaction with different biological structures such as membranes and protein complexes. In these studies, G has been described as a promising vector against bacterial agents as well as a material capable of perturbing biological complexes.

The most important problem in classical simulations of these systems, still now under debate, is the definition of an appropriate set of force fields parameters for G, to allow the implementation of successful simulations with mainstream software packages used for the simulations of biological systems, e.g. GROMACS, (Abraham, 2015), CHARMM (Brooks et al., 2009) or NAMD (Phillips et al., 2005). Although different choices have been investigated, it is commonly accepted to describe G atoms as uncharged Lennard-Jones spheres (Hummer et al., 2001; Patra et al., 2009; Patra et al., 2011). A list of G parameters used in different force fields has been recently reported (Pykal et al., 2016). The aim of this paragraph is to summarize the major findings of computational studies on the interactions between G-based materials and biomolecules, broadly studied at a multi-scale level. A variety of different approaches have been used, such as all-atom or coarse-grained models, and different functional forms and parameters for interactions (the so-called force fields).

A general problem in computational biophysics is the gap between the size and time-scale that can be investigated by simulations and those of biologically relevant mechanisms. Molecular modeling is able to describe biological systems with all-atom details, but this limits its application to study systems of at most ~150 nm and on the microsecond time-scale. A possible way to bridge this gap is to use coarse-grained molecular dynamics (CGMD) simulations, which are based on a controlled reduction of the number of degrees of freedom and the use of shorter-range interacting functions. Due to these simplifications, a CG simulation has a minor resolution but requires less computational resources, allowing the study of larger systems for longer time-scales. A promising approach is to employ a multi-scale description, by alternating the use of classical all-atom with CGMD simulations.

Several computational studies about G-biomolecules systems have been recently published, which can be grouped in the following thematic areas:

1. the adsorption of proteins and peptides (with a particular interest for enzymes and blood proteins) on G substrates, in the context of the study of functional architectures for biomedical applications. Results show that GO has a good solubility in aqueous solution and other organic solvents, thanks to the oxygen-containing groups which can act as reaction sites for the binding of different molecules. As an example, the immobilization of enzymes on a solid substrate is an efficient process to improve its activity while a major factor determining the biocompatibility of a nanomaterial in contact with blood, e.g. medical implants, is the adsorption of proteins on its surface.

2. the interaction of G with biomembranes to assess biological safety or toxicity of G, as well as its promising function as a vector of new classes of antibiotics.

3. DNA or protein detection by G nanopores, an encouraging class of nanosensors that are less sensitive than biological pores to various factors such as the temperature and pH.

In the next paragraph, we describe in more detail some of the studies of points 1 and 2 above, while for those in point 3 we refer the interested reader to the following works for the study of DNA detection (Sathe et al., 2011;Wells et al., 2012;Sathe et al., 2014;Qiu et al., 2015;Barati Farimani et al., 2017) and for the study of protein detection to Farimani *et al.* (Barati Farimani, 2017).

## 5. Adsorption of biomolecules onto Graphene.

### 5.1. Pristine Graphene substrates.

One of the first efforts of all-atom MD simulations for the study of protein adsorption is described in (Zuo, 2011) where the adsorption of the headpiece (HP35) of villin was studied. The simulations showed a rapid adsorption of HP35 by the substrate with relevant conformational changes in both the secondary and tertiary structures. The π/π stacking interactions between aromatic residues and G dominate the protein-G interaction differently from other HP35- curved carbon nanostructures. At a later time, Zhou and collaborators (Gu et al., 2015) performed MD simulations to show how blood proteins such as bovine fibrinogen (BFG) can rapidly adsorb onto the G surface. Markedly, these simulations describe, in addition to the aforementioned effect of strong π/π stacking interactions, another key interaction due to basic residues. These residues play a relevant role during the process because of the strong dispersion interactions between their side-chains and the substrate. Globally, hydrophobic, electrostatic and π/π stacking interactions drive the immobilization of the molecule on G.

In the same year, Kim *et al.* (Kim et al., 2015) examined the recognition of G by peptides with respect to the chemical composition of G, the number of overimposed layers, and the underlying substrate support. The results of this computational work, together with experimental data based on Resonance Raman Spectroscopy, Quartz Crystal Microbalance, and Water Contact Angle measurements, indicate that G quality is a significant factor in G- peptide interactions, while the interaction appears to show no significant dependency on the number of G layers or the underlying support substrate. More recently (No et al., 2017) reported nature-inspired two-dimensional peptide self-assembly on G via optimization of peptide−peptide and peptide−G interactions. Atomistic simulations determined the optimal peptide sequence that leads to peptide self-assembly on G, suggesting that the optimal peptide sequence minimizes the peptide−G interaction energy and also the peptide−peptide interaction energy, resulting in stable complexes on G.

## 6. GO substrates.

The adhesion of biomolecules to GO and rGO layers was investigated in Chong *et al.,* (Chong et al., 2015) to test the different advantages of GO due to the presence of the oxygen atoms. The interactions of serum proteins with GO nanosheets were explored with a large set of experimental techniques and with MD simulations, showing high adsorption capacities of GO and rGO. However, it is important to point out that while GO and rGO were used in the experiments, pristine G was chosen to simulate the relevant non-oxidized regions of the surface present on GO nanosheets. The action of GO was investigated more explicitly by representing the substrate using the Lerf-Klinowski model (Lerf, 1998), which describes the behavior of a standard oxidation process. Using this approach, two paradigmatic papers (Sun et al., 2014; Zeng et al., 2016) demonstrated that GO displays an enhanced adsorption of the attached protein. Firstly, in Sun *et al.* (Sun et al., 2014), an atomistic description of the inhibitory action of GO on the activity of α-chymotrypsin (ChT), has been provided. The results support the hypothesis that GO can be considered as a promising receptor for enzyme inhibition. Secondly, Zeng *et al.* (Zeng et al., 2016) show the details of the binding energy of GO to Vpr13-33, a fragment of the viral protein R (Vpr), using potential of mean force (PMF) calculations with the enhanced method umbrella sampling (Kästner, 2011).

Recently, Willems *et al.* (Willems et al., 2017) investigated the dynamics of supported phospholipid membrane patches stabilized on G surfaces. These systems show potential in sensor device functionalization. The authors integrated experimental measures and CGMD simulations to characterize the molecular properties of supported lipid membranes (SLMs) on G and GO supports. The results described substantial differences in the topologies of the stabilized lipid structures, depending on the nature of the surface, providing novel details into the molecular effects of G and G oxide surfaces on lipid membranes.

Overall, parallel to this considerable amount of data, in this emerging area of computational applications many fundamental issues remain unresolved, due to the lack of sufficient experimental results. In particular, the detailed distribution of the oxygen-containing groups on the substrate is difficult to determine, with a considerable loss in the description of the adsorption mechanism.

## 7. Interactions of Graphene with biomembranes.

The interaction of G with biomolecular complexes is crucial to understand its biological safety and potential toxicity. A seminal work (Tu et al., 2013) showed that GR and GO nanosheets induce the degradation of the inner and outer cell membranes of *Escherichia Coli*. Specifically, MD simulations showed that G is able to actively extract phospholipid molecules from a lipid bilayer, fixing them on its surface. Although these results introduce G as a convenient tool able to kill bacteria, there is an abundant literature where G also shows destructive capacities towards some biomolecules (Luan et al., 2015).

In this context, the results from CGMD simulations describe a quite different scenario. One of the first uses of CGMD for studying G-biomolecules interactions can be found in (Titov et al., 2010). There, the Martini force field (Marrink et al., 2007) is used to study the interaction of G nanosheets with phospholipid bilayers formed by 1-palmitoyl-2-oleoyl-sn-glycero-3-phosphocholine (POPC membrane). Results showed that G sheets are hosted in the hydrophobic interior of the membrane, forming stable G-lipid structures (**Figure 6**).

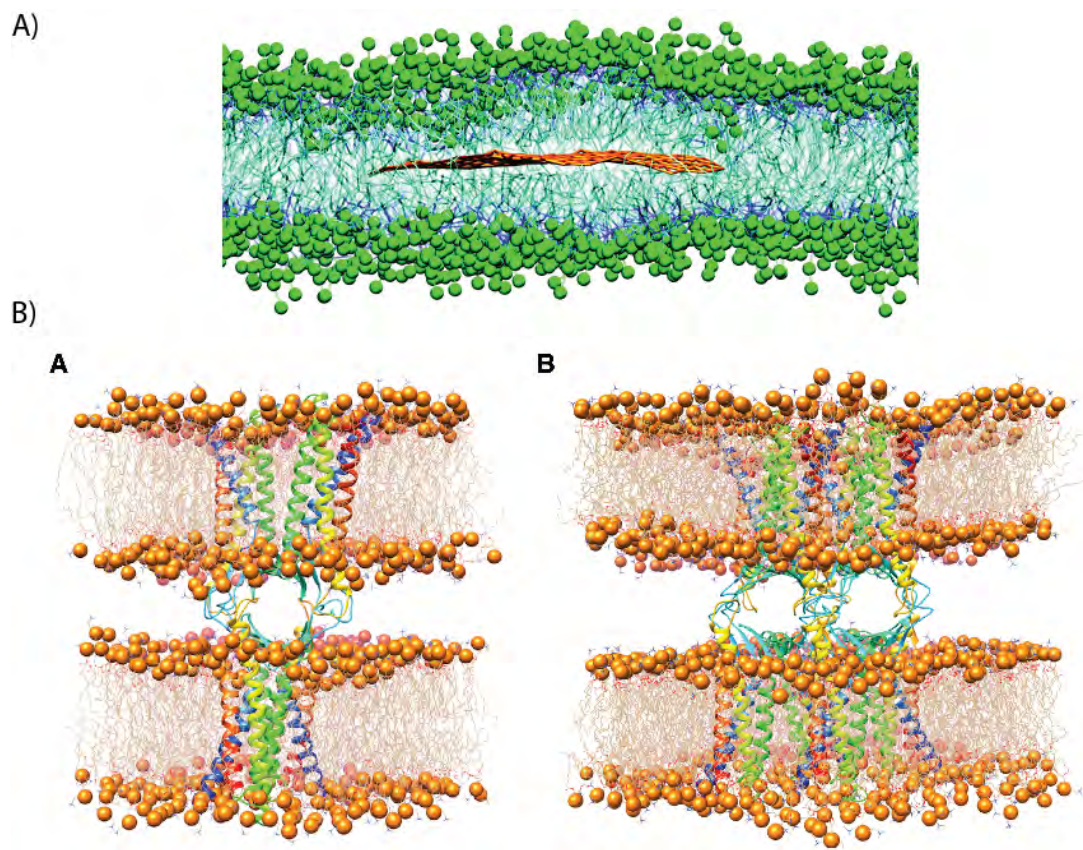

**Figure 6. Graphene interaction with biomembranes.** (A) Equilibrated superstructure of a graphene sheet inside the phospholipid bilayer formed by POPC lipids. Polar heads of the POPC lipids are depicted as green beads, hydrophobic hydrocarbon chains as thick blue lines; the graphene sheet is shown with brown lines (water molecules not shown; modified with permission from Titov et al., (Titov et al., 2010)). (B) The structure of single (**A**) and double (**B**) Cldn15-based paracellular pores, after the respective equilibration protocols. Protomers are shown as ribbons. Each *cis* dimer is embedded in a hexagonal POPC bilayer, shown as wire structures with phosphorus

atoms as spheres. Modified with permission from Alberini et al. (Alberini et al., 2017)

In the following years, other works have investigated these systems with various CGMD algorithms, such as Guo *et al.* (Guo et al., 2013), Mao *et al.* (Mao et al., 2014) and Li *et al.* (Li et al., 2013b). In all these studies, however, lipid extraction or membrane damage is not observed, in contrast to the results of Tu *et al.* (Tu et al., 2013). More recently, computational simulations were used to elucidate whether G causes cell membrane damage (Chen, 2016). All-atom MD simulations were used to study the interaction of both G and GO with respect to a dipalmitoylphosphatidylcholine (DPPC) bilayer, and they revealed that G quickly enters into the membrane by assuming a position parallel to the lipid tails. Conversely, GO did not enter the membrane spontaneously during the observed time-scale, but when docked onto the bilayer, it formed pores in the membrane.

A particularly important biomembrane system is the one involved in the formation of biological barriers such as for example the blood-brain barrier (BBB). Computational studies can be useful to investigate the effect of barrier exposure to G-based materials (**Figure 5**), but studying such complex architectures is still hampered by the lack of structural information (Alberini et al., 2017).

# 8 Conclusions: future challenges and perspectives

In the past few years, GRMs have been studied and used in a wide range of technological fields, including biomedical applications. The treatment of neurological disorders through non-invasive pharmacological approaches is still a major challenge. It is crucial for scientist to develop strategies for efficient cargo delivery of drugs or biomolecules or even genes to the brain, bypassing the BBB while preserving its structure and vital functions. One of the purposes of nanomedicine is indeed to create innovative ways for cell-targeting and drug-controlled release by avoiding surgery or other approaches that are very invasive for the patient. In this scenario, the choice of the appropriate ligand-receptor complex is a key design element when constructing nano-carriers, as well as the choice of the material, the size and eventual functionalization. While receptor-mediated transcytosis is a fundamental pathway for BBB crossing, the development of next-generation nano-carriers, like 2D-materials, and the investigation and optimization of alternative routes for delivery, such as intranasal administration, is of utmost importance for the scientific community.

Besides the 'BBB challenge', other aspects of neuroscience could benefit of the latest developments in graphene research. Neuro-oncology may profit from the development of G nanosheets and G nanoparticles for tumor-targeted imaging, photothermal therapy, and anticancer drug delivery and gene therapy. New electrical, chemical and optical sensors may have great impact for neuro-intensive care and neuro-monitoring. Moreover, the combination of different forms and states of G, diverse chemical functionalization and the possible association with other biomaterials to form G-based composites, may allow to devise an all-in one tool for both diagnosis and therapy, thus effectively building a powerful theranostic device.

Finally, tissue-engineering research is expected to develop novel brain-implant interfaces based on G, to exploit the material electrical conductivity and enhance cell-cell communication and repair. Besides the experimental and clinical evidence, molecular dynamics studies are emerging as an important aspect of material research, as they provide extremely precise indications and predictions on G/cell and G/protein interactions, guiding the researcher to design more powerful G-based devices.

Nevertheless, despite initial studies demonstrated the biocompatibility of G, especially when conjugated with other materials in 2D and 3D scaffolds, only few systems were demonstrated to be successful *in vivo*. Further investigations are still required, in particular about the biological effects of long-term treatment with G materials, before the promised technological applications can be fully exploited in and beyond neuroscience.

**Acknowledgements**
The authors acknowledge financial support from EU H2020 research and innovation programme under grant agreement no. 696656 (Graphene Flaghship – Core1).